Oxi-Shapes: Tropical geometric analysis of bounded redox proteomic state spaces

James N. Cobley

The University of Dundee, Dundee, Scotland, UK.

**Correspondence**: (jcobley001@dundee.ac.uk) or (j_cobley@yahoo.com)

**ORCID**: https://orcid.org/0000-0001-5505-7424

**Abstract**

Redox proteomics generates bounded biochemical measurements that are categorically mismatched to conventional linear algebraic formalisms. This work introduces Oxi-Shapes, a tropical geometric framework for the measurement-native analysis of bounded redox proteomic data. Oxi-Shapes represents cysteine oxidation as a scalar field over a discrete lattice, enabling global and site-wise analysis without rescaling, interpolation, or kinetic assumptions. At the global level, the framework yields internal redox entropy, lattice curvature, and derived energy functionals that characterise the geometric structure of the redox proteome. At the site level, Oxi-Shapes defines a bounded change space that makes explicit hard geometric constraints on admissible redox transitions and enables a normalised signed representation of site-wise change as a fraction of available redox freedom. Applied to an ageing mouse brain dataset, Oxi-Shapes reveals that a small decrease in mean oxidation arises from a profound redistribution of site-wise redox states, with thousands of residues shifting toward the reduced absorbing boundary. These results demonstrate that categorically correct algebraic representations expose structure in proteomic data that is inaccessible to mean-centric or unbounded analyses.

## Introduction

Chemically reversible cysteine oxidation redox regulates biological processes by post-translationally modifying protein structure and function[1]. Advances in redox proteomics have enabled large-scale quantification of cysteine oxidation across thousands of sites, tissues, and conditions, providing detailed snapshots of redox state across the proteome[2–4]. However, the analytical frameworks commonly used to interpret these data often rely on algebraic assumptions that are mismatched to the structure of the underlying biochemical state space.

Cysteine oxidation measurements are intrinsically bounded to the unit interval and defined over discrete, invariant residue identities[5]. They report biochemical state occupancies rather than unconstrained continuous variation. Bounded biochemical state spaces with absorbing boundaries and many-to-one observation are categorically incompatible with linear and Euclidean representations. In such spaces, translation invariance, global invertibility, and symmetry do not hold, and equal-magnitude numerical changes do not encode equal physical meaning[6–8]. Consequently, common operations such as linear projection, Euclidean distance, or variance maximisation can generate invalid states, obscure boundary-induced structure, and irretrievably conflate distinct biochemical histories. The consequence of this algebraic mismatch is a loss of access to physically meaningful information about the order–disorder, entropy, and energy of the measured redox states.

This work introduces Oxi-Shapes, a tropical geometry-based framework for analysing redox proteomic data that is categorically matched to bounded biochemical state spaces. In Oxi-Shapes, invariant cysteine sites define a discrete lattice, and measured oxidation occupancies are represented as a bounded scalar field over this lattice. Structure is therefore induced directly by experimental measurements rather than imposed by an external embedding. Within this representation, entropy corresponds to geometric volume, order–disorder emerges naturally along the bounded oxidation axis, and curvature and energetic structure can be computed using well-defined Dirichlet and Morse formulations, with fully reduced states constituting the ground configuration[9]. Without invoking dynamics or mechanisms, Oxi-Shapes establishes what can be physically inferred from bounded biochemical snapshots once representation is correct. The Oxi-Shapes framework can be applied to any bounded biochemical state space that is categorically matched to its algebra, including other post-translational modifications such as protein phosphorylation[10].

## Results
### Overview
The analyses presented in this work proceed from a fixed mathematical representation of bounded biochemical state spaces. The algebraic structure and geometric properties of the Oxi-Shapes representation are derived formally in the Supplemental Notes, independent of a given empirical dataset. Implementation details, including data preprocessing and construction of the discrete lattice representation, are described in Methods and applied here to a foundational redox proteomic dataset (Oxi-Mouse[11]).

The results are organised around geometric objects that exist on bounded biochemical state spaces rather than around analytical procedures. These objects fall into two complementary classes. Local geometric objects describe site-wise structure and admissible state transitions within the bounded oxidation interval, capturing capacity, feasibility, and directionality of redox change at individual residues. Global geometric objects describe high-dimensional organisation of the redox state across the proteome, capturing ensemble-level order–disorder, entropy, and energy.

For each object, this work establishes what can and cannot be inferred under many-to-one observation and then apply it to experimental redox proteomic data. This structure ensures that each result reflects a necessary consequence of categorically correct representation rather than an outcome of optimisation, modelling, or algorithmic choice.

**The cysteine redox proteome as a high-dimensional discrete tropical lattice**

As a measurement-native framework, Oxi-Shapes operates on datasets in which cysteine oxidation is quantified as bounded occupancies in the $[0,1]$ state space. To represent these data as a high-dimensional object, each cysteine residue is assigned a unique and invariant position in a discrete lattice index set, providing a fixed correspondence between measured sites and lattice elements across conditions. Measured oxidation values are then realised as bounded scalar values along a third (oxidation) axis over this lattice (**Supplemental Note 1**), yielding a discrete tropical representation in which structure is induced directly by the measured biochemical state rather than by imposed relational or embedding assumptions.

In Oxi-Shapes, each cysteine site therefore carries an intrinsic state variable—its oxidation occupancy—that is mutable across conditions but conserved at the level of information: site identity is invariant, and each measurement assigns a well-defined bounded value whenever the site is observed. This property enables direct information-theoretic[12] interpretation of the measured redox states (i.e., non-probabilistic) without logarithmic transforms, binning procedures, or assumptions of unbounded state space.

To make this representation concrete, an example instantiation of the cysteine redox proteome—computed in young and old mouse brain[11]—in which each invariant lattice element corresponds to a measured cysteine site and the bounded oxidation occupancy is shown as a scalar height (**Fig. 1**). The brain was selected due to its sensitivity to redox dysregulation[13–15]. This example serves solely to illustrate the measurement-native lattice object populated by experimental data; no relational structure or geometric operators are introduced at this stage. As a measurement-native structure, the lattice object does not encode biological embeddings, such as grouping of indexes by pathway terms[16–18].

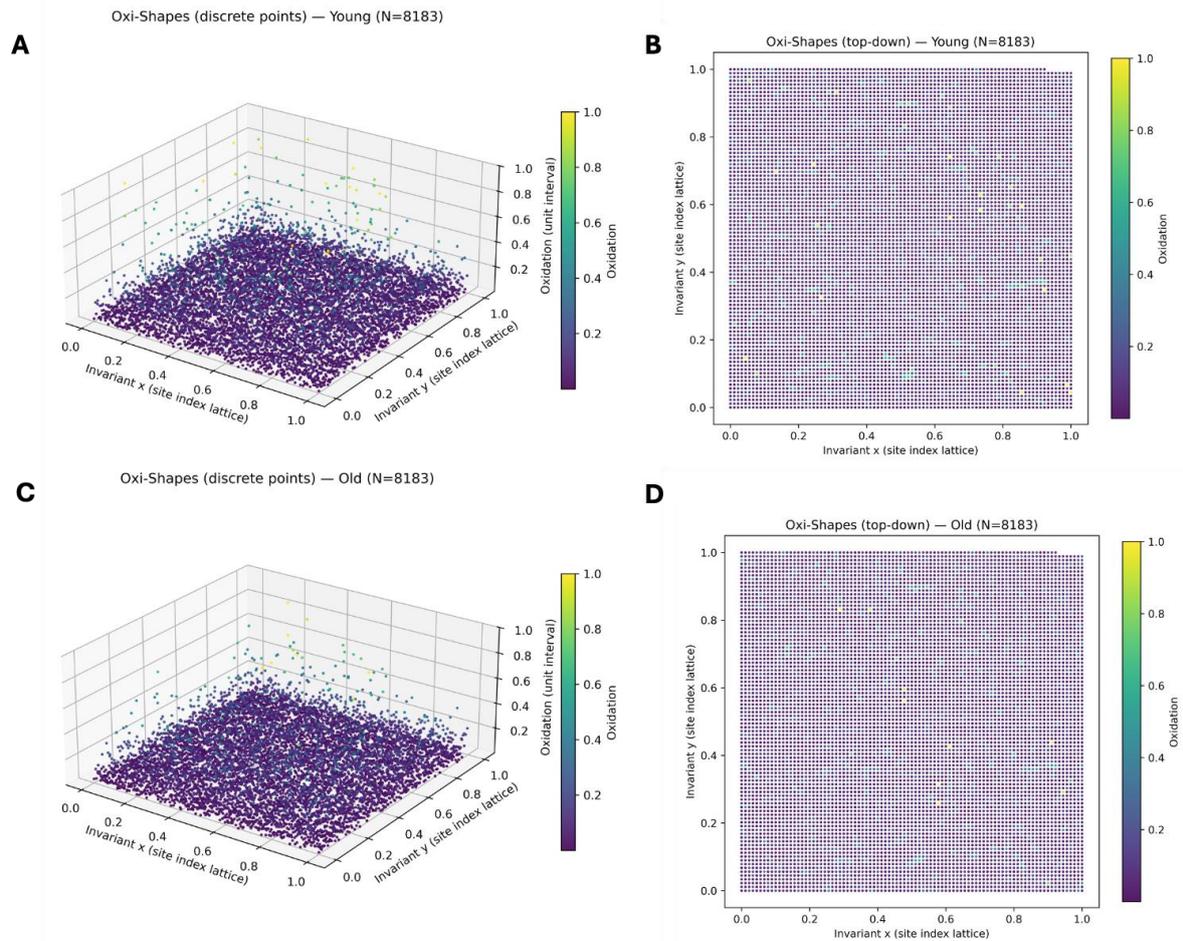

**Figure 1 | The cysteine redox proteome as a discrete tropical lattice.** Each point corresponds to an invariant cysteine site positioned on a discrete lattice index set. Colour and height indicate the measured oxidation occupancy bounded to the unit interval [0,1]. Top-down (B,D) and three-dimensional views (A,C) are shown for the same object. Young (A,B) and old (C,D) mouse brain datasets are displayed as example instantiations of the lattice populated by experimental measurements.

## Internal redox entropy: order–disorder in the bounded cysteine state space

In Oxi-Shapes, internal redox entropy defines a volume-based measure of order–disorder on the discrete cysteine lattice (**Supplemental Note 2**). By construction, each site contributes exactly one bounded scalar value in [0,1], and the total volume of the object is given by the arithmetic mean across sites. This quantity corresponds to the average uplift of the lattice along the bounded oxidation axis, providing a direct geometric measure of how evenly reduced and oxidised states are distributed across the admissible state space (**Fig. 1**).

Maximum internal order corresponds to every site occupying either the 0 (reducing bias) or 1 (oxidising bias) state, which geometrically maps to a perfectly flat lattice at 0 (zero volume) or 1 (maximal volume). In a bounded binary state space, maximum internal disorder occurs at 0.5 occupancy, corresponding to the region of greatest configurational degeneracy. While no dynamical behaviour is assumed or measured here, this point represents the most disordered macrostate in the admissible configuration space, from which no additional disorder can be gained without altering system constraints.

Internal redox entropy is not a measure of thermodynamic entropy and makes no direct claim about heat exchange, dissipation, or microscopic irreversibility[19–21]. The measured configurations instead reflect these properties insofar as they arise from an open biological system, but they are not themselves measurements of external entropy production. Rather, internal redox entropy represents a geometric projection of configurational structure defined on a bounded biochemical state space at a fixed observational snapshot. Consistent with a tropical representation, this construction is inherently mesoscopic and non-invertible: multiple biochemical histories may map to the same observed configuration[5], while dynamical irreversibility remains outside the scope of the present analysis.

Applying this framework to shared cysteine sites quantified in young and old mouse brain ($N$ = 8,183 sites) provides a proof-of-principle instantiation of internal redox entropy in a biological system rather than a test of a statistical hypothesis. Within this framework, ageing was associated with a decrease in internal redox entropy of the cysteine proteome, reflected as a reduction in mean oxidation from young to old brain (Young = 0.0889; Old = 0.0814; Δ = −0.0074).

Although the absolute change in mean oxidation is numerically small, its interpretation must be made relative to the bounded state space. The young proteome begins at 0.0889, meaning the maximal possible shift toward a fully reduced state is itself only 0.0889. The observed decrease therefore represents a substantial fraction of the available reducing headroom. Both young and old proteomes occupy a highly ordered regime near the saturating 0 boundary, where additional reductions in mean correspond to progressively stronger constraints on admissible configurations.

Consistent with this geometry, internal redox entropy was numerically identical to the arithmetic mean in both conditions, confirming that the volume measure introduces no additional transformations beyond those inherent to the bounded representation. However, when mapped onto the combinatorial structure of the binary state space[22], the observed shift corresponded to a pronounced reduction in configurational degeneracy. Despite similar mean oxidation, the young and old proteomes differed by approximately 63 orders of magnitude in the number of admissible configurations consistent with their respective means (**Fig. 2**).

The discrete lattice visualisation captures the specific realised configuration in each condition, while the degeneracy curve quantifies the size of the surrounding admissible configuration space indexed by the same mean. Together, these results demonstrate that ageing of the mouse brain is associated with a measurable contraction of the accessible redox configuration space, even when changes in mean oxidation appear modest. Internal redox entropy thus provides a geometrically grounded descriptor that distinguishes realised redox organisation from the broader space of mathematically admissible alternatives.

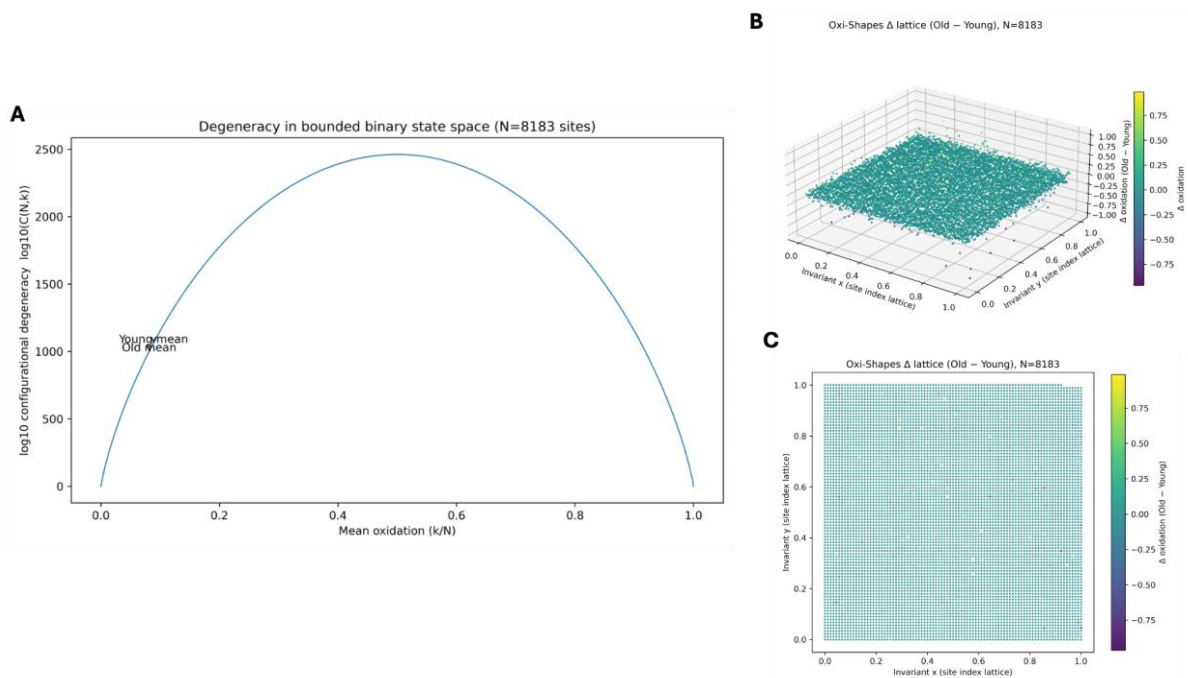

**Figure 2 | Internal redox entropy and configurational degeneracy in the ageing mouse brain
(A)** Configurational degeneracy in a bounded binary state space for *N* = 8,183 cysteine sites. The number of admissible configurations consistent with a given mean oxidation (k/N) is given by the binomial coefficient Ω = C (N, k), shown here on a $\log_{10}$ scale. The resulting inverted-U curve reflects maximal degeneracy at 0.5 occupancy and minimal degeneracy at the boundaries. Vertical markers indicate the observed mean oxidation values for young and old mouse brain, both of which lie deep within the low-occupancy, highly ordered regime.
**(B)** Three-dimensional Oxi-Shapes Δ-lattice showing site-wise changes in oxidation between old and young mouse brain (Old − Young). Each point corresponds to a single cysteine site positioned on the invariant lattice; colour and height encode the signed change in oxidation within the bounded interval [−1, 1]. Despite the modest shift in the global mean, the lattice reveals a structured redistribution of oxidation states across sites.
**(C)** Top-down view of the same Δ-lattice, emphasising the spatial distribution of redox changes without imposing adjacency or continuity. Together, panels (B) and (C) visualise the realised configuration corresponding to the observed change in internal redox entropy, complementing the abstract degeneracy structure shown in (A).

### Deriving an energy functional from the bounded redox lattice scalar field

In Oxi-Shapes, the primary measurement object is a bounded scalar field $\phi \in [0,1]^N$ defined on a discrete index lattice. Volume-based order–disorder metrics are therefore definable without imposing any neighbourhood structure (**Supplemental Notes 1–2**). However, any energy functional that depends on spatial variation—i.e., that penalises or rewards changes in $\phi$ across the lattice—requires an explicit relational operator.

This work therefore introduces a minimal, index-defined adjacency on the lattice and use the associated graph Laplacian to derive curvature-like and energy-like functionals from the same measurement-native field, enabling high-dimensional geometric analysis of the lattice object. It does not encode physical interaction networks, spatial proximity, reaction pathways, or temporal transitions between cysteine residues. Neighbour relations are introduced solely to enable geometric comparison of bounded state values and carry no biological interpretation beyond this formal role. Importantly, these constructions do not claim thermodynamic energy measurement; they define mathematically well-posed functionals on a bounded biochemical state space that enable consistent comparisons across conditions (**Supplemental Note 3**).

Applying a minimal index-defined adjacency enabled the computation of curvature- and energy-like functionals directly from the measured redox lattice without altering the underlying scalar field. The resulting graph Laplacian curvature exhibited zero mean in both young and old proteomes, as required by construction (**Fig. 3A**), but showed a systematic reduction in dispersion with age (Young SD = 0.437; Old SD = 0.376). This narrowing of curvature fluctuations indicates a reduction in local geometric variation across the lattice, consistent with a loss of configurational heterogeneity in the aged proteome despite similar global ordering. Importantly, curvature here reflects discrete second-order variation of the measured field and does not imply spatial diffusion, interaction, or dynamics.

From this relational structure, the Dirichlet energy, which quantifies quadratic variation of oxidation values across the lattice, was defined. Dirchlet energy was substantially lower in old relative to young brain ($\Delta E\_D$ = −41.3 total; −26% per-node mean), indicating reduced spatial roughness of the redox field with age (**Fig. 3B**). Hence, the topology shifted toward a "flatter" state, indicative of less volume (**Fig. 1-2**).

Likewise, a Morse-type potential with a ground state at $\varphi = 0$ yielded lower mean and total energy in the old proteome ($\Delta$ mean = −0.012; $\Delta$ sum = −98.6), reflecting a global shift toward lower energetic displacement from the reducing ground state (**Fig. 3C**). Put differently, the morse energy reflects the volume deflating toward the ground state—0, indicating a contraction toward an absorbing 0 state.

Without involving thermodynamics or kinetics, these results demonstrate that ageing is associated with a coordinated reduction in curvature and energy-like variation across the bounded biochemical state space.

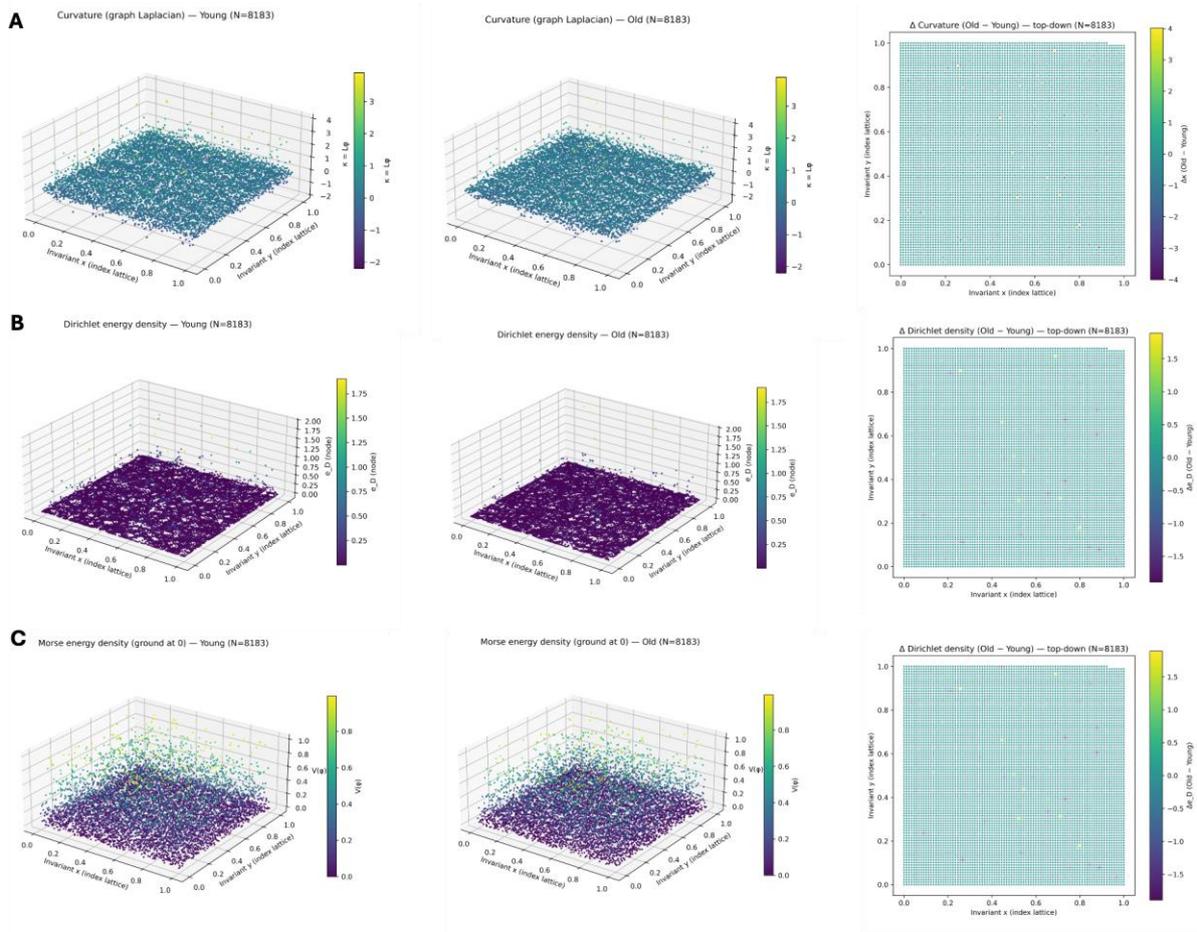

**Figure 3 | Curvature and energy functionals derived from the bounded redox lattice scalar field.**
**(A)** Graph Laplacian curvature (κ = Lφ) computed on the discrete index lattice for young (left) and old (middle) mouse brain proteomes (N = 8,183 shared cysteine sites), visualised as three-dimensional Oxi-Shapes with lattice indices on the horizontal axes and curvature on the vertical axis. The corresponding top-down difference map (right) shows Δκ = κ_old − κ_young. By construction, curvature has zero mean and captures second-order variation of the bounded scalar field.
**(B)** Dirichlet energy functional density per node, quantifying quadratic variation of oxidation values across adjacent lattice indices, shown for young (left) and old (middle) conditions, with the top-down difference map (right) indicating a global reduction in Dirichlet energy with age.
**(C)** Morse energy functional density with a ground state at φ = 0, shown for young (left) and old (middle) proteomes, and corresponding top-down difference map (right). This measurement native functional assigns increasing energy cost to displacement from the fully reduced state while remaining defined entirely on the bounded scalar field.

## Site-wise bounded tropical geometry of redox state changes

Having characterised the global geometric structure of the redox proteome, this work examines the geometry of site-wise redox state changes within the bounded cysteine state space (**Supplemental Note 4**). For each cysteine residue, the measured oxidation in the initial condition defines a starting point $x_A \in [0,1]$, while the change between conditions is given by $\Delta = x_B - x_A$. Each site is therefore represented as a point $(x_A, \Delta)$ in a constrained two-dimensional space whose admissible region is fully determined by the boundedness of oxidation occupancy.

This representation is exact and measurement-native. No rescaling, normalisation, or linear projection is applied, and the feasible region $-x_A \leq \Delta \leq 1 - x_A$ follows directly from the unit interval constraint. The geometry of this site-wise change space therefore encodes the full set of physically admissible redox transitions without imposing assumptions about dynamics, kinetics, or biological coupling between residues.

Because the global redox state is defined as the arithmetic mean of site-wise occupancies, any difference between global means must arise exclusively from sites that change state. Sites with $\Delta \approx 0$ contribute identically to both conditions and therefore cannot influence the global shift. Consequently, the observed change in global redox organisation is necessarily encoded entirely in the subset of sites exhibiting non-zero signed change. This property follows directly from the linearity of the mean and is independent of any assumptions about mechanism or regulation.

This observation enables a natural geometric decomposition of the lattice into symmetric and asymmetric components. Sites with $|\Delta|$ below a specified tolerance define a symmetric subspace that preserves the global redox state across conditions, while sites with signed changes define an asymmetric subspace that drives the observed shift. The balance between these components provides a site-resolved encoding of global redox symmetry breaking without invoking statistical modelling or parameter fitting.

Site-wise redox change is strictly bounded by the starting occupancy of each site. For a site with initial oxidation $x_A$, admissible changes satisfy $-x_A \leq \Delta \leq 1 - x_A$. This constraint implies that although each site formally has one degree of freedom, any realised change represents a signed collapse constrained by its available headroom. For example, a site at $x_A = 0.4$ may increase by at most 0.6 or decrease by at most 0.4. If a decrease occurs, no further reduction is possible without violating the bounded state space. These hard geometric limits impose strict constraints on physical action at the level of individual residues and make explicit how identical numerical changes can carry fundamentally different physical meaning depending on starting state[5].

To enable comparison of site-wise changes across different starting occupancies, a normalised signed redox change that expresses each realised $\Delta$ as a fraction of the site's available redox freedom was defined. For a site with initial oxidation $x_A$, the normalised change is defined as $a = \Delta/(1 - x_A)$ for oxidation and $a = \Delta/x_A$ for reduction, yielding a bounded coordinate $a \in [-1,1]$. This representation preserves directionality while removing dependence on absolute headroom, providing a constraint-respecting, representation-only descriptor of site-wise redox change.

Applying this site-wise bounded representation to shared cysteine sites quantified in young and old mouse brain (N = 8,183), the mean oxidation decreased modestly with age (Young = 0.0889; Old = 0.0814; $\Delta$ = −0.00742). Consistent with the linearity of the mean, this global shift was driven exclusively by sites exhibiting non-zero signed change. Using a tolerance of $|\Delta| \leq$ 0.001 to define identity, 217 sites (2.7%) were classified as symmetric, while 7,966 sites (97.3%) formed the asymmetric subspace encoding the entire global redox difference. Reconstructing the global mean change from asymmetric sites alone reproduced the observed $\Delta$ to numerical precision.

Site-wise changes occupied the full admissible region of the bounded change space, with Δ spanning from −0.963 to +0.988 depending on starting oxidation. When expressed as a normalised signed redox change, many sites exhibited values approaching |a| ≈ 1, indicating near-complete utilisation of available redox freedom despite modest absolute changes. Across asymmetric sites, the distribution of a was skewed toward negative values (mean a = −0.223), consistent with the observed global shift toward a more reduced proteome. These results demonstrate that the bounded tropical geometry yields a faithful, constraint-respecting decomposition of global redox change into site-resolved contributions without invoking statistical modelling or dynamical assumptions.

Despite exhibiting similar global mean oxidation, the young and old redox lattices are almost completely divergent in their site-wise composition. The small net shift in the mean is produced by a large redistribution of individual cysteine states, with thousands of sites undergoing signed, bounded transitions rather than remaining invariant. Notably, a substantial subset of sites exhausts nearly all available redox freedom, collapsing toward the fully reduced (0) absorbing boundary, revealing extensive geometric reorganisation beneath an apparently modest global change.

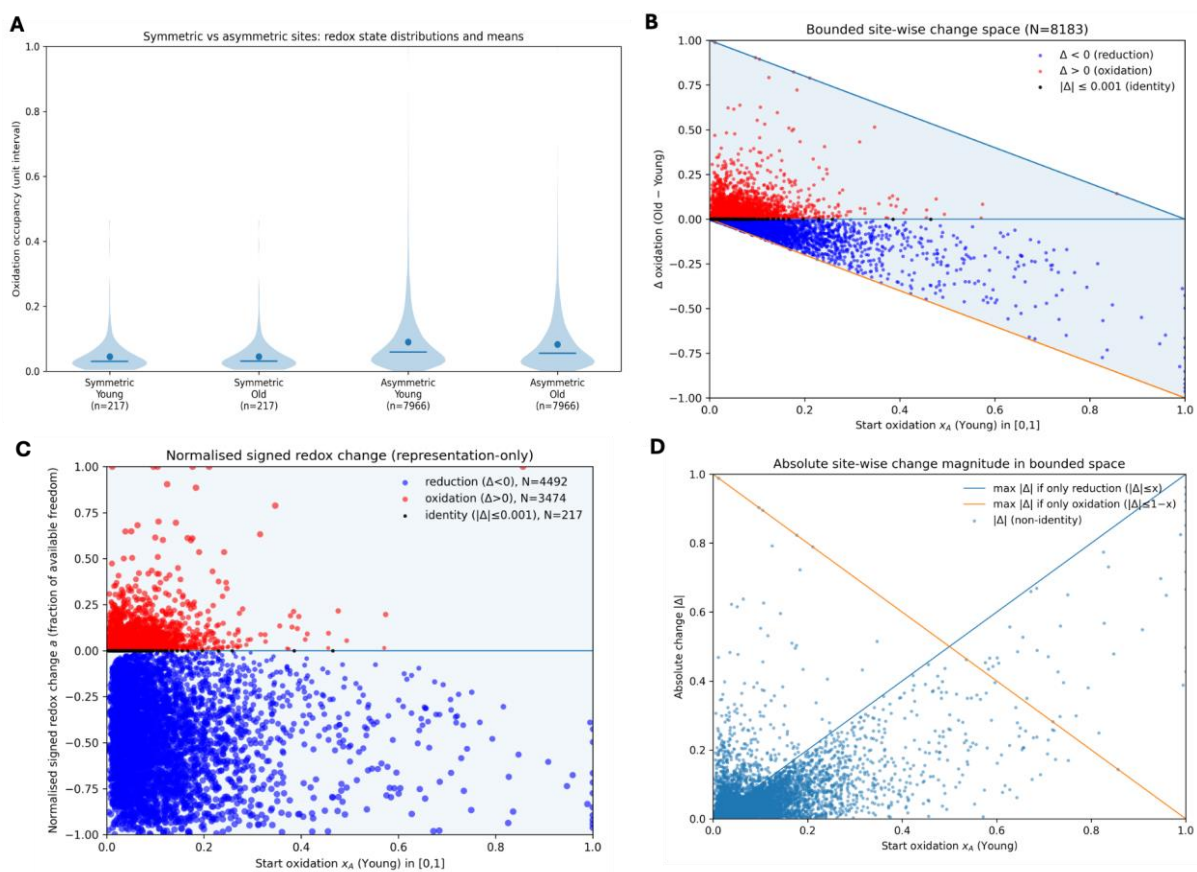

**Figure 4 | Site-wise bounded tropical geometry of redox state changes.** **(A)** Violin plots showing the distribution of oxidation occupancies for symmetric sites ($|\Delta| \leq 0.001$) and asymmetric sites ($|\Delta| > 0.001$) in young and old mouse brain. Points and horizontal bars indicate means. Symmetric sites contribute identically to both conditions and therefore do not influence the global redox shift, whereas asymmetric sites encode the entirety of the observed change. **(B)** Exact site-wise change space represented as starting oxidation $x_A$(young) versus signed change $\Delta = x_B - x_A$. The admissible region is strictly bounded by $-x_A \leq \Delta \leq 1 - x_A$, shown by the triangular envelope. Points are coloured by direction of change (reduction, oxidation, or identity), illustrating the hard geometric constraints

imposed by the bounded state space.
**(C)** Normalised signed redox change $a$, defined as the signed fraction of available headroom used by each site. This representation collapses site-wise changes onto a common scale $[-1,1]$ while preserving directionality, making explicit how identical absolute changes correspond to different fractions of available freedom depending on starting state.
**(D)** Absolute site-wise change magnitude $|\Delta|$ as a function of starting oxidation, shown together with the maximal admissible change envelopes for pure reduction ($|\Delta| \leq x_A$) and pure oxidation ($|\Delta| \leq 1 - x_A$). This panel highlights how large absolute changes are geometrically restricted near the boundaries of the bounded redox space.

### Discussion

Motivated by the goal of representing and analysing empirical data using a categorically correct algebraic formalism, the present work introduced Oxi-Shapes as a tropical geometric framework for analysing bounded redox proteomic data. At its core, Oxi-Shapes applies measurement-native tropical algebraic constructions to high-dimensional biochemical measurements, operating both at the level of the global object—the discrete lattice—and at the level of site-wise state changes. This dual perspective enables the computation of internal redox entropy, curvature, and derived energy-like functionals from the same bounded measurement space, while also exposing strict geometric constraints on admissible site-wise redox changes.

Computational instantiation of these constructions demonstrates the value of the framework. Application to the Oxi-Mouse dataset revealed that ageing is associated with a decrease in the internal redox entropy of the cysteine proteome, corresponding to a substantial reduction in the configurational degeneracy available to the global lattice. This shift was accompanied by a flattening of the lattice topology, reflected in reduced curvature, and by decreases in both Dirichlet and Morse energy functionals. Together, these changes indicate that ageing moves an already highly ordered manifold closer to the absorbing 0 boundary of the bounded state space.

Mathematically, these global shifts arise from a profound reorganisation of the cysteine proteome. Thousands of sites altered their redox occupancy, with a non-trivial fraction exhausting their available redox headroom and collapsing fully toward the reducing boundary. These behaviours reveal hard geometric constraints on permissible directional changes given a site's starting oxidation state, which are made explicit by the tropical geometric representations introduced here. Notably, this extensive redistribution occurred despite only a marginal change in the arithmetic mean oxidation, demonstrating that closely similar means can correspond to radically different underlying configurations—consistent with the high degeneracy of the bounded state space at low occupancy.

This proof-of-principle application shows that Oxi-Shapes can be used to test experimental hypotheses without introducing additional modelling assumptions. Although demonstrated here using cysteine redox data, the underlying mathematics is agnostic to residue type and is directly applicable to other bounded biochemical measurements, such as stoichiometric methionine oxidation or tyrosine phosphorylation.

Beyond formal algebraic correctness, Oxi-Shapes motivates a conceptual shift in how proteomic data are interpreted. While quantities such as internal redox entropy and Morse energy do not measure physical entropy or thermodynamic energy, they reveal the shadow of biochemical action on the proteome. In particular, the Morse energy reflects the constraint that many sites occupy or approach an absorbing reduced state, which can only be escaped through real biochemical work. More broadly, the framework makes explicit an epistemic limit of inference: snapshot measurements collapse many non-equivalent biochemical trajectories into a single observed configuration, and dynamics cannot be uniquely reconstructed from such data alone.

## Methods
### Redox proteomics dataset
Redox proteomics data were obtained from the Oxi-Mouse dataset[11], in which cysteine oxidation was quantified as percentage occupancy for individual cysteine residues across biological conditions. For the analyses shown here, oxidation measurements from young and old mouse brain samples were used as representative instantiations of the redox state space. Oxidation values were expressed as percentages or fractions depending on source and were converted to unit interval representation $[0,1]$ prior to analysis. All downstream analyses were performed on this bounded state space.

### Preprocessing and site selection
To ensure that comparisons were performed on invariant molecular identities, analyses were restricted to cysteine sites measured in both conditions under consideration. Sites with missing oxidation values in either condition were excluded. Each cysteine site was assigned a stable and deterministic identifier derived from available metadata (e.g., protein accession, residue position, peptide sequence). When no explicit site identifier was available, the row index was used as a fallback. This ensured that each site retained a unique and invariant identity across conditions and analyses. Oxidation values were converted to the unit interval $[0,1]$.

### Discrete lattice construction
To construct the Oxi-Shapes object, each invariant cysteine site was embedded in a discrete two-dimensional lattice index space. Site identifiers were first sorted deterministically, and lattice coordinates were assigned sequentially on a near-square grid. The resulting coordinates were normalised to $[0,1]$ for visual consistency across datasets of different sizes. This lattice embedding is index-based only and serves to provide a discrete base manifold for representing bounded state measurements. It does not encode biological adjacency, interaction, similarity, or spatial proximity between cysteine residues.

For each condition, the measured oxidation occupancy of each cysteine site was represented as a bounded scalar value assigned to its corresponding lattice position. The resulting object is a high-dimensional discrete lattice with a scalar field defined over it, fully determined by experimental measurements. The object therefore reflects only invariant site identity and bounded biochemical state, consistent with the formal construction (**Supplemental Note 1**).

Oxi-Shapes were visualised using both top-down and three-dimensional scatter plots. In the three-dimensional view, lattice indices define the horizontal axes and oxidation occupancy defines the vertical axis. In the top-down view, oxidation occupancy is encoded by colour. All visualisations were rendered as discrete point clouds without interpolation, triangulation, or smoothing, to avoid implicitly imposing continuity or neighbourhood structure. Colour scales and axis limits were shared across conditions to ensure direct comparability.

**Computation of internal redox entropy and configurational degeneracy**

Internal redox entropy was computed directly from the bounded cysteine oxidation measurements. For a given condition, internal redox entropy was defined as the arithmetic mean of site-wise oxidation occupancies across all invariant cysteine sites retained for analysis. Because each site contributes exactly one bounded scalar value in $[0,1]$, this quantity is numerically identical to the total lattice volume scaled by the number of sites, consistent with the formal definition of the Oxi-Shapes object (**Supplementary Notes 1–2**).

To verify numerical consistency, internal redox entropy was explicitly compared to the arithmetic mean oxidation for each condition, and equivalence was confirmed to machine precision. The difference in internal redox entropy between conditions was reported as the difference in mean oxidation.

Configurational degeneracy associated with a given internal redox entropy value was computed under a bounded binary-state approximation. For a system of $N$ cysteine sites and a measured mean oxidation $V$, the effective number of oxidised sites was defined as $k = \text{round}(N \times V)$. The number of distinct configurations consistent with this mean was then given by the binomial coefficient

$$\Omega(N, k) = \binom{N}{k}.$$

Degeneracy values were reported on a logarithmic scale due to their large magnitude.

To visualise the structure of the admissible configuration space, degeneracy was computed across the full bounded range of mean oxidation values from 0 to 1 in fixed increments, yielding a characteristic inverted-U curve with maximal degeneracy at 0.5 occupancy. Observed mean oxidation values for each biological condition were mapped onto this curve to contextualise the realised configurations within the surrounding admissible configuration space.

This analysis is deterministic and proof-of-principle in nature. No statistical hypothesis testing was performed. Internal redox entropy and configurational degeneracy were computed solely to characterise the geometric and combinatorial structure of the bounded cysteine state space defined by experimental data.

**Computation of lattice curvature and derived energy functionals**

To derive energy-like descriptors from the bounded redox lattice, a minimal relational structure was introduced on the discrete index grid. A 4-neighbour adjacency (von Neumann neighbourhood) was imposed on the two-dimensional lattice used for Oxi-Shapes visualisation, connecting each lattice site to its immediate horizontal and vertical neighbours where present. This adjacency is index-defined only and does not encode biological interaction networks, spatial proximity, reaction pathways, or temporal transitions between cysteine residues.

The resulting adjacency structure defines a graph $G = (V, E)$ with $|V| = N$ nodes corresponding to invariant cysteine sites and edges connecting neighbouring lattice indices. From this graph, the unnormalised graph Laplacian was constructed as

$$L = D - A,$$

where $A$ is the adjacency matrix and $D$ is the diagonal degree matrix.

Laplacian curvature was computed by applying the graph Laplacian to the bounded oxidation field,

$$\kappa = L\phi,$$

yielding a discrete second-difference operator that quantifies local variation of oxidation values relative to neighbouring lattice indices. Curvature values were summarised by their mean and standard deviation and visualised both as three-dimensional lattice scatter plots and as top-down difference maps between conditions.

A quadratic variation energy functional (Dirichlet energy) was computed as

$$E_D(\phi) = \frac{1}{2} \sum_{(i,j) \in E} (\phi_i - \phi_j)^2 = \frac{1}{2} \phi^\top L \phi.$$

Total Dirichlet energy was reported for each condition, along with a per-node energy density obtained by normalising by the number of sites. This quantity measures the aggregate spatial variation of oxidation values across the lattice and depends only on relative differences between adjacent lattice indices.

To encode an explicit ground-state bias at the fully reduced state ($\phi = 0$), a bounded Morse-type potential was applied independently to each lattice site,

$$V(\phi_i) = (1 - e^{-a\phi_i})^2,$$

with parameter $a = 6.0$. Morse energy was summarised as both the mean and total energy across all sites. This functional depends only on the absolute oxidation value at each site and does not require neighbourhood interactions.

**Construction of the bounded site-wise change space**

Site-wise redox state changes were analysed using a bounded tropical geometric representation derived directly from experimental measurements. For each invariant cysteine site measured in both conditions, the oxidation occupancy in the initial condition (Young) was denoted as

$$x_A \in [0,1],$$

and the change between conditions was defined as

$$\Delta = x_B - x_A,$$

where $x_B$ is the oxidation occupancy in the comparison condition (Old).

Each cysteine site was therefore represented as a point $(x_A, \Delta)$ in a two-dimensional bounded change space. Because oxidation occupancy is constrained to the unit interval, admissible changes are strictly bounded by

$$-x_A \leq \Delta \leq 1 - x_A.$$

This feasible region arises directly from the boundedness of the underlying biochemical measurement and was not imposed algorithmically. Sites violating these bounds were explicitly checked for and none were observed beyond numerical tolerance.

This representation is measurement-native and exact. No rescaling, transformation, regression, or dimensionality reduction was applied. The geometry of the site-wise change space therefore encodes the complete set of physically admissible redox transitions consistent with the experimental data.

**Classification of symmetric and asymmetric sites**

To decompose the site-wise change space, cysteine residues were classified according to the magnitude of their signed change. Sites with

$$|\Delta| \leq \varepsilon$$

were classified as *symmetric* (identity-preserving), where $\varepsilon = 10^{-3}$ corresponds to a tolerance of 0.1% oxidation occupancy. Sites with $|\Delta| > \varepsilon$ were classified as *asymmetric* and represent residues contributing to differences between conditions.

This classification follows directly from the linearity of the arithmetic mean: sites with $\Delta \approx 0$ contribute identically to both conditions and therefore cannot influence the global mean oxidation, whereas sites with non-zero signed change necessarily encode all observed differences between global states. No statistical modelling or hypothesis testing was applied to define these classes.

**Normalised signed redox change (fraction of available freedom)**

To enable comparison of site-wise changes across different starting occupancies, a normalised signed redox change coordinate was defined. For each asymmetric site, the realised change $\Delta$ was expressed as a fraction of the site's available redox headroom,

$$a = \begin{cases} \dfrac{\Delta}{1 - x_A}, & \text{if } \Delta > 0 \, (\text{oxidation}) \\[2mm] \dfrac{\Delta}{x_A}, & \text{if } \Delta < 0 \, (\text{reduction}) \\[2mm] 0, & \text{if } |\Delta| \leq \varepsilon. \end{cases}$$

This yields a bounded, dimensionless coordinate $a \in [-1, 1]$, where the sign encodes directionality and the magnitude encodes the fraction of available redox state space utilised by the observed change. Values of $a = \pm 1$ correspond to sites that fully exhaust their admissible redox headroom in the observed direction.

This normalised signed redox change is a representation-only descriptor. It preserves directionality while removing dependence on absolute starting occupancy and does not assume kinetic, temporal, or mechanistic equivalence between sites.

**Ranking of site-wise action magnitude**

To provide an ordering of site-wise redox changes independent of absolute scale, the absolute normalised change $|a|$ was ranked among asymmetric sites using a dense rank normalised to the unit interval. This *delta-rank* provides a representation-only measure of relative action magnitude within the bounded change space. Rank values were used exclusively for visual encoding and descriptive reporting and were not interpreted as probabilities or statistical significance measures.

**Visualisation of site-wise redox geometry**
Four complementary visualisations were generated:

1. **Bounded site-wise change space**: scatter plot of $(x_A, \Delta)$ with explicit rendering of the admissible region $-x_A \le \Delta \le 1 - x_A$, coloured by directionality (oxidation, reduction, identity).
2. **Absolute change magnitude**: plot of $|\Delta|$ as a function of $x_A$, overlaid with the theoretical upper bounds imposed by oxidation and reduction headroom.
3. **Normalised signed redox change**: scatter plot of $(x_A, a)$, with point size optionally encoding delta-rank.
4. **Symmetric vs asymmetric distributions**: violin plots comparing oxidation occupancy distributions in Young and Old conditions for symmetric and asymmetric site classes, with arithmetic means overlaid.
5. 

All plots were rendered directly from discrete data points without interpolation, smoothing, or kernel density estimation beyond standard violin plot construction. Axis limits and colour encodings were fixed across conditions to ensure direct comparability.

**Deterministic decomposition of global mean change**
As a consistency check, the contribution of asymmetric sites to the global mean oxidation difference was computed explicitly. Because symmetric sites satisfy $\Delta \approx 0$ by construction, the global mean change can be reconstructed as the mean $\Delta$ among asymmetric sites weighted by their fraction of the total site count. Agreement with the directly computed global mean difference confirms that all observed changes in global redox state arise exclusively from asymmetric site-wise contributions. This decomposition is deterministic and follows directly from the linearity of the arithmetic mean.

**Data and code availability**
The redox proteomics data were downloaded from a previously published and publicly available proteomics resource deposited in a community-accessible website (https://oximouse.hms.harvard.edu/). No new primary data were generated.

All original code used to construct and visualise Oxi-Shapes is openly available at https://github.com/JamesCobley/Oxi-Shapes/tree/main. An archived, immutable version of the code corresponding to this manuscript has been released https://github.com/JamesCobley/Oxi-Shapes/releases.

**Acknowledgements.**

The author thanks Prof. Angus I. Lamond (the University of Dundee) and all of the members of the Lamond lab for constructive scientific discussions. The support of an MCR grant (MR/Y013891/1) is gratefully acknowledged.

**Conflict of interest**
The author declares that there are no conflicts of interest.

**Declaration of generative AI and AI-assisted technologies in the writing process**
During the preparation of this work, the author used ChatGPT (OpenAI) as an AI-assisted tool to support language editing, structural refinement, and clarification of conceptual arguments. All scientific content, theoretical development, interpretations, and conclusions were conceived by the author. The author reviewed, edited, and verified all text generated with AI assistance and takes full responsibility for the content of the published article.

# Supplemental Notes

# Supplemental Note 1 | Mathematical construction of Oxi-Shapes

### Bounded biochemical state space

Let $\mathcal{L} = \{1, \ldots, N\}$ denote a finite index set corresponding to invariant molecular identities (here, cysteine residues). For each site $i \in \mathcal{L}$, an experimental measurement assigns a scalar oxidation occupancy

$$x_i \in [0,1],$$

representing the fraction of molecules observed in an oxidised state at that site. The unit interval constitutes a **bounded biochemical state space** with absorbing boundaries at 0 (fully reduced) and 1 (fully oxidised).

### Measurement-native object construction

An Oxi-Shape is defined as a mapping

$$f: \mathcal{L} \to [0,1], i \mapsto x_i,$$

assigning a bounded scalar state to each invariant site identity.

To visualise this object, sites are embedded in a discrete lattice index space by assigning each $i \in \mathcal{L}$ a unique and invariant pair of coordinates $(u_i, v_i)$, forming a two-dimensional discrete lattice. This embedding is **index-based only** and does not encode biological proximity, similarity, or interaction.

The measured oxidation value $x_i$ is represented as a bounded scalar deformation along a third axis. The resulting object is a **high-dimensional discrete lattice with a bounded scalar field**, fully determined by experimental measurements.

At this stage, the object carries **no topology, adjacency, metric, or relational structure** beyond site identity and bounded state assignment.

### Algebraic constraints of bounded state spaces

Because $x_i \in [0,1]$, the state space is not closed under unconstrained linear operations. In particular:

- Translation invariance does not hold.
- Linear combinations can generate invalid states outside $[0, 1]$.

- Equal numerical differences do not correspond to equal physical changes near boundaries.
- Global invertibility and symmetry assumptions underlying Euclidean vector spaces are violated.

As a consequence, applying linear projections, Euclidean distances, or variance-maximising operators to bounded biochemical occupancies is **categorically incompatible** with the data-generating process.

Oxi-Shapes therefore adopts a **bounded (tropical) algebraic perspective**, in which admissible operations must respect the absorbing boundaries and non-linearity intrinsic to the state space.

## 4 Information conservation and state occupancy

In Oxi-Shapes, each site $i$ carries a state variable $x_i$ that is **mutable across conditions** yet **conserved at the level of information**: site identity is invariant, and whenever a site is observed it is assigned a well-defined bounded value.

There are no null internal states within the measurement space; absence of information arises only from missing observations, not from the algebra of the state space itself.

**Corollary.**
This construction enables direct information-theoretic interpretation of biochemical state without logarithmic transforms, binning procedures, or assumptions of unboundedness. Measures of order, disorder, and entropy may therefore be defined directly on the bounded object once a representation consistent with the algebra is established.

## S1.5 Scope and non-assumptions

The construction above is intentionally minimal. At this stage, Oxi-Shapes does **not** assume:

- temporal dynamics or trajectories,
- kinetic or mechanistic models,
- neighbourhood relations between sites,
- smoothness or continuity,
- biological interaction networks.

All subsequent geometric, energetic, or relational structures are introduced only after this measurement-native object has been defined.

# Supplementary Note 2 | Volume and configurational degeneracy in bounded cysteine state space

## Definition of the bounded cysteine state space

Consider a protein system with $N$ cysteine sites, each of which can occupy a reduced or oxidised state. In Oxi-Shapes, each site $i$ is associated with a bounded scalar oxidation value

$$x_i \in [0,1],$$

representing the fraction of molecules in which that cysteine is oxidised at the time of measurement.

The admissible biochemical state space is therefore bounded and discrete, with each site contributing exactly one scalar degree of freedom. The full configuration of the system at a given observational snapshot is represented by the vector

$$\mathbf{x} = (x_1, x_2, \ldots, x_N).$$

## Volume as a geometric invariant

The total volume $V$ of an Oxi-Shapes object is defined as the arithmetic mean of site-wise oxidation values:

$$V \;=\; \frac{1}{N}\sum_{i=1}^{N} x_i\,.$$

Geometrically, this quantity corresponds to the average uplift of the discrete lattice along the bounded oxidation axis. Because each site contributes exactly one bounded scalar, the volume is invariant under lattice permutation and independent of site ordering.

Importantly, no transformation, normalisation, or rescaling is applied beyond the intrinsic unit interval already encoded in the data. By construction, this volume is numerically identical to the arithmetic mean oxidation, but it is here interpreted as a geometric invariant of the lattice.

## Internal redox entropy as an order–disorder coordinate

We define **internal redox entropy** as this volume coordinate $V$. This quantity provides a measure of order–disorder on the bounded cysteine lattice:

- **Maximum internal order** occurs when all sites occupy the same boundary state (all 0 or all 1), corresponding to a perfectly flat lattice with minimal configurational freedom.
- **Maximum internal disorder** occurs at intermediate occupancy, where reduced and oxidised states are most evenly distributed.

Crucially, this definition makes no claim about thermodynamic entropy, heat exchange, dissipation, or microscopic irreversibility. Internal redox entropy is a *geometric projection* of configurational structure defined on a bounded biochemical state space at a fixed observational snapshot.

## Configurational degeneracy of the mean

While the volume *V* uniquely specifies the mean oxidation, it does **not** uniquely specify the underlying configuration. For a system of *N* binary sites, the number of distinct configurations consistent with a given mean oxidation is given by the binomial coefficient:

$$\Omega(N, k) = \binom{N}{k},$$

where

$$k = N \times V.$$

This quantity measures the **configurational degeneracy** associated with a given volume coordinate. As *V* varies from 0 to 1, the degeneracy exhibits a characteristic inverted-U structure, with:

- minimal degeneracy at the boundaries (*V* = 0 or 1),
- maximal degeneracy at *V* = 0.5.

This structure reflects a fundamental property of bounded binary state spaces and is independent of any biological assumptions.

## Interpretation of realised configurations

The observed Oxi-Shapes lattice corresponds to one realised configuration among the $\Omega$ admissible configurations consistent with the measured mean oxidation. The lattice visualisation therefore specifies **which configuration occurred**, while the degeneracy curve quantifies **how many alternative configurations were mathematically possible but not realised**.

In this sense, internal redox entropy decomposes naturally into:

- a **real component** (the volume / mean oxidation), and
- a **configurational context** (the degeneracy associated with that mean).

This separation clarifies how small changes in mean oxidation can correspond to disproportionately large changes in the size of the admissible configuration space, particularly near the boundaries of the bounded state space.

## Scope and limitations

This construction is explicitly mesoscopic and non-invertible. Multiple biochemical histories may map to the same observed configuration, and no claims are made regarding the dynamics, kinetics, or external entropy production that gave rise to the measured state.

Rather, Supplemental Note 2 establishes a mathematically complete and bounded framework for interpreting cysteine redox measurements as geometric objects with well-defined volume and configurational structure.

# Supplemental Note 3 | Laplacian curvature and derived energy functionals on the bounded lattice

Let $\phi \in [0,1]^N$ denote the bounded occupancy field defined on a discrete index lattice of $N$ sites. A minimal 4-neighbour adjacency is imposed on the lattice, yielding a graph $G = (V, E)$ with $|V| = N$. Let $A$ denote the adjacency matrix and $D$ the diagonal degree matrix. The (unnormalised) graph Laplacian is defined as

$$L = D - A.$$

We define **Laplacian curvature** on the lattice as the action of the graph Laplacian on the bounded field,

$$\kappa \equiv L\phi,$$

which constitutes a discrete second-difference operator measuring local variation of the scalar field relative to its neighbours.

A quadratic variation **energy functional** (Dirichlet energy) is defined by

$$E_D(\phi) \equiv \frac{1}{2} \sum_{(i,j) \in E} (\phi_i - \phi_j)^2 = \frac{1}{2} \phi^\top L \phi.$$

This quantity measures the total squared variation of the field across adjacent lattice indices and may be expressed as a per-node energy density by distributing edge contributions equally to incident nodes.

To encode an explicit ground-state bias at $\phi = 0$, we define a bounded Morse-type potential at each lattice site, parameterised by $a > 0$,

$$V(\phi_i) = (1 - e^{-a\phi_i})^2.$$

The resulting **Morse energy functional** is summarised either as the mean

$$\frac{1}{N}\sum_{i=1}^{N} V(\phi_i)$$

or as its total sum over nodes.

All functionals defined here are derived directly from the measurement-native bounded field $\phi$ and depend only on the chosen index adjacency. They are introduced as mathematically well-posed descriptors of variation and energy on a bounded biochemical state space. These quantities do **not** represent thermodynamic energy, heat exchange, dissipation, or microscopic irreversibility.

# Supplemental Note 4 | Site-wise bounded tropical geometry of redox state changes

Let $\phi \in [0,1]^N$ denote the vector of site-wise cysteine oxidation occupancies measured for a given condition, where each component corresponds to a single cysteine residue. For two conditions $A$ and $B$ (e.g. young and old), define the site-wise change field

$$\Delta = \phi^{(B)} - \phi^{(A)}.$$

For each site $i$, the initial occupancy $x_{A,i} = \phi_i^{(A)}$ defines a starting point in the bounded unit interval, and the pair $(x_{A,i}, \Delta_i)$ uniquely represents the site-wise redox transition between conditions.

**Bounded site-wise change space**

Because oxidation occupancy is bounded to the unit interval, admissible changes satisfy the hard constraints

$$-x_{A,i} \leq \Delta_i \leq 1 - x_{A,i}.$$

Hence, all site-wise transitions lie within a closed, triangular region of the $(x_A, \Delta)$ plane. This feasible region is determined entirely by the bounded state space and does not depend on assumptions about kinetics, dynamics, or regulation.

This construction is measurement-native: no rescaling, normalisation, or projection is applied. The resulting geometry therefore encodes the complete set of physically admissible site-wise redox transitions directly implied by the data.

**Symmetric and asymmetric site decomposition**

The global redox state of a sample is defined as the arithmetic mean of site-wise occupancies,

$$\bar{\phi} = \frac{1}{N} \sum_{i=1}^{N} \phi_i \,.$$

By linearity, the difference between global means across conditions is given by

$$\bar{\phi}^{(B)} - \bar{\phi}^{(A)} = \frac{1}{N} \sum_{i=1}^{N} \Delta_i \,.$$

Sites with $\Delta_i = 0$ contribute identically to both conditions and therefore cannot influence the global shift. Consequently, any non-zero global change must arise exclusively from sites exhibiting signed change.

This motivates a geometric decomposition of the lattice into:

- a **symmetric subspace**, defined by sites with $\mid \Delta_i \mid \leq \varepsilon$, and
- an **asymmetric subspace**, defined by sites with $\mid \Delta_i \mid > \varepsilon$,

where $\varepsilon$ is a fixed tolerance reflecting measurement precision. This decomposition is exact and parameter-free aside from the explicit tolerance choice.

**Headroom-limited action and signed collapse**

Although each site formally has one degree of freedom, any realised change is constrained by its available headroom. For a site with initial oxidation $x_A$, the maximum admissible oxidation is $1 - x_A$, while the maximum admissible reduction is $x_A$.

Thus, a realised change represents a signed collapse constrained by these limits. For example, a site at $x_A = 0.4$ may increase by at most $0.6$ or decrease by at most $0.4$; if a decrease occurs, no further reduction is possible without violating the bounded state space. These constraints impose strict geometric limits on site-wise physical action.

**Normalised signed redox change**

To enable comparison of site-wise changes across different starting occupancies, we define a **normalised signed redox change**

$$a_i = \begin{cases} \dfrac{\Delta_i}{1 - x_{A,i}}, & \Delta_i > 0 \\[2mm] \dfrac{\Delta_i}{x_{A,i}}, & \Delta_i < 0 \\[2mm] 0, & \mid \Delta_i \mid \leq \varepsilon, \end{cases}$$

which yields a bounded coordinate $a_i \in [-1,1]$.

This representation preserves directionality while expressing each realised change as a fraction of the site's available redox freedom. Importantly, $a$ is a representation-only quantity: it does not encode dynamics, rates, or energetic costs, but provides a constraint-respecting geometric descriptor of site-wise redox change.

**Summary**

Together, the bounded change space $(x_A, \Delta)$, the symmetric/asymmetric decomposition, and the normalised signed coordinate $a$ provide a complete geometric description of site-wise redox state transitions implied by bounded occupancy measurements. These constructions depend only on the unit-interval constraint and arithmetic structure of the data and therefore constitute an exact, assumption-free geometric encoding of site-wise redox reorganisation.